\documentclass[12pt]{article}

\usepackage{graphicx}

\newcommand{\be}{\begin{equation}}
\newcommand{\ee}{\end{equation}}
\newcommand{\bea}{\begin{eqnarray}}
\newcommand{\eea}{\end{eqnarray}}

\newcommand{\tA}{\tilde A}
\newcommand{\tB}{\tilde B}

\newcommand{\RR}{\rangle}
\newcommand{\LL}{\langle}

\makeatletter
\newdimen\normalarrayskip
\newdimen\minarrayskip
\normalarrayskip\baselineskip
\minarrayskip\jot
\newif\ifold             \oldtrue

\makeatother
\newlength{\extraspace}
\newlength{\extraspaces}
\input{tcilatex}

\begin{document}

\addtolength{\baselineskip}{.4mm}

\thispagestyle{empty}

\begin{flushright}
\baselineskip=12pt
$         $\\
\hfill{  }\\
\end{flushright}
\vspace{.5cm}

\begin{center}
\baselineskip=24pt

{\LARGE {Universal Manipulation of a Single Qubit}}\\[15mm]

\baselineskip=12pt

{Lucien Hardy and David D. Song} \\[%
8mm]
{ Centre for Quantum Computation\\[0pt]
Clarendon Laboratory, Department of Physics \\
University of Oxford, Parks Road, Oxford OX1 3PU, U.K.} \vspace{2cm}

{\bf Abstract}

\begin{minipage}{15cm}
\baselineskip-12pt

We find the optimal universal way of manipulating a single qubit,
$|\psi (\theta,\varphi)\RR$, such that $(\theta, \varphi) \rightarrow
(\theta-k, \varphi - l)$.
Such optimal transformations fall into two classes.  For $0\leq k \leq
\pi/2$ the optimal map is the identity and the fidelity varies
monotonically from 1 (for $k=0$) to $1/2$ (for $k=\pi/2$).  For
$\pi/2 \leq k \leq \pi$ the optimal map is the universal-NOT gate and
the fidelity varies monotonically from $1/2$ (for $k=\pi/2$) to $2/3$
(for $k=\pi$).
The fidelity $2/3$ is equal to the fidelity of
measurement.  It is therefore rather surprising that for some values of $k$
the fidelity is lower than $2/3$.

\end{minipage}
\end{center}
\vspace{2cm}
A unit of classical information is a bit, i.e. 0 or 1.
 Quantum information consists of qubits which are superposition of the states
 $|0\RR$ and $|1\RR$.
 Classical and quantum information differs in many ways.
   While classical information can be copied perfectly, 
the same is not true with qubits  \cite{zurek,clon1,clon2}.
   Another feature that distinguishes classical and quantum information is a measurement.
 Unlike classical information, an unknown single qubit cannot be
 measured to give complete information about the qubit.
In order to get a maximum information about an unknown qubit,
we measure the qubit along any chosen basis $\{ |\phi\RR , |\phi^{\perp}\RR\}$. 
If the result is $|\phi\RR$, then we guess the unknown qubit to be $|\phi\RR$ and if
the result is $|\phi^{\perp}\RR$, then we guess $|\phi^{\perp}\RR$.
Averaging over all possible $|\phi\RR$'s (assuming a uniform
distribution over the Bloch sphere), the fidelity is equal to 2/3.  We
cannot achieve a higher fidelity by using generalised measurements and
hence 2/3 is the optimal measurement fidelity of an unknown qubit.

A qubit in Bloch vector notation is
\be
|\psi(\theta,\varphi)\RR \equiv \left( \begin{array}{c}
\cos(\theta/2) \\ e^{-i\varphi } \sin(\theta /2)
\end{array} \right)
\ee
The most general linear transformation on $(\theta,\varphi)$ is
\be (\theta,\varphi) \rightarrow (\theta-k,\varphi-l) \label{general}\ee
with $0\leq k \leq \pi$ and $0\leq l \leq 2\pi$. If $\theta=0,\pi$ then
$\varphi$ is undefined.  For definiteness we will take $\varphi=0$ in such
cases.  Since, when taking averages over the Bloch sphere, these
anomalous cases are of measure zero we need not pay any special attention to them.
This general transformation can be composed from two transformations.  First
\be (\theta,\varphi) \rightarrow (\theta-k,\varphi) \label{thetatrans}\ee
and then $\varphi \rightarrow \varphi -l$.  The transformation on $\varphi$ can
be achieved perfectly by a unitary operation and so is of less interest
to us.  However, the transformation on $\theta$ cannot be achieved
unitarily.  To find the fidelity of general linear transformations of
the form (\ref{general}) it suffices to consider only the non-unitary part
(\ref{thetatrans}).  We are interested only in universal
transformations. These are those transformations for which the fidelity
is independent of $\theta$ and $\varphi$ of the input state.
Furthermore, we will assume that the input distribution is uniform over
the Bloch sphere.  Since the area element $\sin\theta d\theta d\varphi$
is not preserved in form (except when $\theta=0,\pi$) the output will
not be uniform.  In taking averages we integrate over a uniform
distribution of the input variables which corresponds to integrating
over a non-uniform distribution of the output variables.

Changing bits 0 to 1 and 1 to 0 is a NOT-gate in classical information case.
 In quantum case, changing $|\psi\RR=a|0\RR+b|1\RR$ to $|\psi^{\perp}\RR =
 b^* |0\RR-a^* |1\RR$ requires anti-unitary transformation which is not allowed
 in quantum mechanics.  In \cite{buzek1,buzek2,gisin}, it was shown that
 universal-NOT(U-NOT) operation can be achieved with 2/3 fidelity for a
 single input. This fidelity is same as the measurement fidelity.
They showed that the U-NOT operation is no better than measuring a qubit first then
preparing an orthogonal state. In Bloch vector notation
the  U-NOT gate corresponds to transforming
$|\psi(\theta,\varphi)\RR$ to $|\psi(\theta-\pi,\varphi)\RR$.
This is a special case of the transformation (\ref{thetatrans}) with
$k=\pi$.  Now consider the general case in which we transform
$|\psi(\theta,\varphi)\RR$ to $|\psi(\theta -k,\varphi)\RR$.
Naively, it may seem that the fidelity should
be at least 2/3, since one could measure a qubit with $2/3$ fidelity and prepare a  state
in an appropriate direction.  We will show in this paper that this is not so.

Let us take an example where $k=3\pi/4$. Therefore for a given unknown state
$|\psi\RR$, we want to prepare a state as close as possible to
$|\psi^{\prime}\RR =|\psi(\theta-3\pi/4,\varphi)\RR $.
We choose a random state
\be
|\phi(u,v)\RR \equiv
\left( \begin{array}{c}
\cos(u/2) \\ e^{-iv } \sin(u/2)
\end{array} \right)
\ee
and measure $|\psi(\theta,\varphi)\RR$ in the basis of
$\{ |\phi\RR,|\phi^{\perp}\RR \}$. If we get $|\phi\RR$ we prepare
$|\phi^{\prime}\RR\equiv |\phi(u-3\pi/4,v)$ and if we get $|\phi^{\perp}\RR$, then
$|\phi^{\prime\perp}\RR$ is prepared.  As a density matrix, the state
we prepare by this method is
\be
\rho^{(1)} = |\LL \psi|\phi\RR|^2 |\phi^{\prime} \RR\LL \phi^{\prime}| + |\LL \psi|\phi^{\perp}\RR|^2 |\phi^{\prime\perp}\RR\LL \phi^{\prime\perp}|
\ee
We take the average of $\rho^{(1)}$ over uniform distributions of
$|\phi\RR$ on the Bloch sphere to obtain $\overline{\rho^{(1)}}$ and then
the fidelity is given by
\be
F = \frac{1}{4\pi} \int_0^{\pi}\int_0^{2\pi} \LL
\psi^{\prime}|\overline{\rho^{(1)}}|\psi^{\prime}\RR \sin \theta d\theta d\varphi = .5833...
\label{F0}\ee
This value is lower than 2/3.  Can we do any better? 
 If we prepare $|\phi^{\perp}\RR$ when the result is $|\phi\RR$ and prepare
$|\phi\RR$ for $|\phi^{\perp}\RR$, i.e.
\be
\rho^{(2)} =   |\LL \psi|\phi\RR|^2 |\phi^{\perp} \RR\LL \phi^{\perp}| + |\LL \psi|\phi^{\perp}\RR|^2 |\phi\RR\LL \phi |
\ee
then
\be
F=  \frac{1}{4\pi} \int_0^{\pi}\int_0^{2\pi} \LL
\psi^{\prime}|\overline{\rho^{(2)}}|\psi^{\prime}\RR \sin \theta d\theta d\varphi = .6178...
\ee
This fidelity is still lower than 2/3 but higher than the value in (\ref{F0}). 
This rather surprising result is due to the
different phase angles of $|\psi\RR$ and $|\phi\RR$.
Suppose $|\LL \psi|\phi\RR|^2 = 2/3$, then the rotation $|\psi(\theta,\varphi)\RR \rightarrow
|\psi(\theta-\pi,\varphi)\RR$ and $|\phi(u,v)\RR\rightarrow |\phi(u-\pi,v\RR$ yields the same
fidelity $|\LL \psi(\theta-\pi,\varphi)|\phi(u-\pi,v)\RR|^2 = 2/3$.
However if the rotation is some other angle $k\neq \pi$ or  $0$, then
$|\LL \psi(\theta-k,\varphi)|\phi(u-k,v)\RR|^2$ may not be 2/3 because $\varphi$ and $v$ are not necessarily the same. 
If the phase angles $\varphi$ and $v$ are the same, then for any $k$,  $|\LL \psi(\theta-k,\varphi)|\phi(u-k,\varphi)\RR|^2 = 2/3$.
For $\pi/2 \leq k \leq \pi$ in $|\psi(\theta-k,\varphi)\RR$,
$\overline{\rho^{(2)}}$ yields the fidelity
\be 
F^{(2)} = \frac{1}{12}(6 + \cos (\pi-k) + \cos (\pi +k))
\label{F1}\ee
For $0\leq k\leq\pi/2$, we consider the usual measurement density matrix,  
\be
\rho^{(3)} =   |\LL \psi|\phi\RR|^2 |\phi \RR\LL \phi| + |\LL \psi|\phi^{\perp}\RR|^2 |\phi^{\perp}\RR\LL \phi^{\perp} |
\ee
and the fidelity is given as 
\be
F^{(3)}=\frac{1}{2} + \frac{1}{6} \cos k
\label{F3}\ee
For $\pi/2 \leq k \leq \pi$, we will show that $F^{(2)}$ in (\ref{F1}) is indeed the optimal
fidelity. 
$|\psi(\theta-k,\varphi)\RR$ for $0\leq k\leq \pi/2$ can be
obtained with better fidelity than $F^{(3)}$ in (\ref{F3}).
We expect this since for $k=0$, the identity operation gives $|\psi\RR$ with fidelity 1.

By considering the most general type  of transformation on a single
qubit we will find the one which maximises the fidelity for the
transformation (\ref{thetatrans}).  We will follow the method of
Buzek {\it et al.} \cite{buzek2}.  The most general operation available
to us is to
perform unitary evolution on the single qubit and some ancilla prepared
in a known state $|Q\RR$ (this is taken to be normalised). This gives
\bea
|0\RR|Q\RR &\rightarrow & |1\RR|A\RR + |0\RR |B\RR \nonumber \\
|1\RR|Q\RR &\rightarrow & |0\RR|\tA\RR + |1\RR|\tB\RR \label{Transf}
\eea
where $|A\RR,|\tA\RR,|B\RR,|\tB\RR$ may not be normalised. 
 From the normalisation and the orthogonality of
(\ref{Transf}), $|A|^2 + |B|^2 = |\tA|^2+|\tB|^2 = 1$ and $\LL A|\tB\RR+ \LL B|\tA\RR =0$.
 We let $|\psi\RR$ transform under (\ref{Transf}), trace over the ancilla
 and obtain a density matrix $\rho^{({\rm out})}$ for our qubit.
The fidelity is then given by $F=\LL \psi^{\prime}|\rho^{({\rm out})}|\psi^{\prime}\RR$
 where $|\psi^{\prime}\RR = |\psi(\theta-k,\varphi)\RR$.  When expressed
explicitly in terms of $\theta$, $\varphi$, $|A\RR$, $|B\RR$, $|\tA\RR$,
and $|\tB\RR$ this expression for the fidelity has 48 terms.
We can compare coefficients of those terms dependent on $e^{\pm i\varphi}$
and $e^{\pm 2i\varphi}$.  These coefficients must vanish in order for the transformation in (\ref{Transf}) to be
independent of $\varphi$ (which we require for universality),
$  \LL A|\tA\RR = \LL A|B\RR$ = $\LL \tB|\tA\RR = \LL \tA|B\RR$=$\LL \tB |A\RR=0$.
Of those terms remaining, some have a dependence on $\theta$.  These
terms must also vanish (by universality).  This leaves only two terms
giving us an expression for the fidelity:
\be
F=\sin^2\frac{k}{2}|\tA|^2 + \cos^2 \frac{k}{2} |\tB|^2
  = (\cos^2 \frac{k}{2} -\sin^2\frac{k}{2} )|\tB|^2 + \sin^2 \frac{k}{2}
\label{fidelity}\ee
By comparing coefficients of functions of $\theta$ of those terms having
a $\theta$ dependence and setting them to zero we obtain the
following conditions
\bea
 & & 2|\tA|^2-2|\tB|^2 + \LL \tB|B\RR+\LL B|\tB\RR = 0 \label{i} \\
 & & 2|B|^2 - 2|A|^2 - \LL \tB|B\RR - \LL B|\tB\RR = 0 \label{ii} \\
 & & |A|^2 + |\tA|^2 -|B|^2- |\tB|^2 + \LL B|\tB\RR + \LL \tB|B\RR =0 \label{iii} \\
 & & (\cos^2\frac{k}{2}-2\sin^2\frac{k}{2})|\tA|^2 + (\sin^2\frac{k}{2} -
 2\cos^2\frac{k}{2})|\tB|^2+\sin^2\frac{k}{2} |B|^2 + \cos^2\frac{k}{2}|A|^2 \nonumber \\
  & &+(\cos^2\frac{k}{2}-\sin^2\frac{k}{2})\LL \tB|B\RR + (\cos^2\frac{k}{2}-\sin^2\frac{k}{2})\LL B|\tB\RR =0  \label{iv} 
\eea
\begin{figure}[ptb]
\begin{center}
{\includegraphics[scale=1.0]{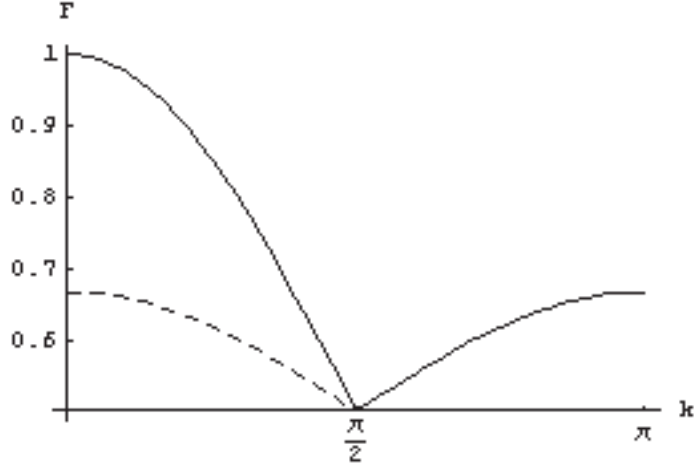}}
\end{center}
\caption{A graph of $F$ versus $k$ is shown.  For $0 \leq k \leq \pi/2$,
the upper curve corresponds to the optimal quantum scheme
and the lower curve represents the measurement scheme.
For $\pi/2 \leq k \leq \pi$, both measurement and optimal quantum schemes yield the
identical results. }
\end{figure}
From (\ref{i}) and (\ref{ii}),
\be
|A|^2 = |\tA|^2 \;\;\; , \;\;\; |B|^2 = |\tB|^2
\label{v}\ee
which then implies (\ref{iv}) is equal to (\ref{ii}). From (\ref{iii}), with $\eta = {\rm Re}(\LL B|\tB\RR)/|\tB|^2$ (therefore $|\eta|\leq 1$),
\be
|\tB |^2 = \frac{1}{2-\eta}
\ee
For $\pi/2 \leq k \leq \pi$, $|\tB|^2$ needs to be minimum to give a maximum fidelity in (\ref{fidelity}).
Therefore with $\eta=-1$, the fidelity is
\be
F=\frac{1}{3}\cos^2 \frac{k}{2} + \frac{2}{3} \sin^2 \frac{k}{2}
\label{F5}\ee
which is same as (\ref{F1}).  Therefore for $\pi/2 \leq k \leq \pi$, the measurement based preparation as in $\rho^{(2)}$ 
is indeed optimal.  The transformation satisfying (\ref{i}-\ref{v}) and (\ref{F5}) is
same as the U-NOT transformation of Buzek {\it et al.} in \cite{buzek1,buzek2}.
The fidelity in (\ref{F5}) has the highest value of 2/3 when $k=\pi$ and the lowest 1/2 when $k=\pi/2$.  
The graph for $k$ and $F$ is shown in Figure 1 where the measurement and the quantum schemes yield the identical result.
For $0\leq k\leq \pi/2$, $|\tB|^2$ needs to be maximum to have a maximum fidelity in (\ref{fidelity}).  
Therefore with the choice of $\eta=1$, 
\be
F=\cos^2 \frac{k}{2}
\ee
This transformation is simply a trivial identity map and it has maximum fidelity of 1 when $k=0$ and minimum 1/2 when 
$k=\pi/2$.  A graph of $k$ and $F$ for $0\leq k\leq \pi/2$ is shown in
Figure 1. In this case, the quantum scheme has higher fidelity than the
measurement scheme.

It follows that for a general transformation linear in the spherical
coordinates, namely $(\theta,\varphi)\rightarrow (\theta-k,\varphi-l)$, the
procedures which optimise fidelity fall into two distinct classes.  (A) For
$0\leq k\leq \pi/2$ the optimal procedure is the identity map which
performs better than a measurement based scheme.  In this range the
maximum fidelity (equal to one) is achieved, not surprisingly,  when $k=0$.
(B) For $\pi/2 \leq k \leq \pi$ the U-NOT transformation is optimal.
This procedure performs only as well as a measurement based scheme.  In
this range the maximum fidelity (equal to $2/3$) is achieved, perhaps a
little surprisingly, only for the case $k=\pi$ which, if $l=0$, corresponds to
a universal NOT operation. Since $\varphi$ can be varied linearly by a
unitary transformation, $l$ can take any value in either of these two
classes.

\parindent=0pt
\vspace{6mm}
{\bf Acknowledgements}
\vspace{6mm}

\parindent=6mm
           
We are grateful to Leah Henderson and Vlatko Vedral for discussions on
this topic.  LH is funded by a Royal Society University Research Fellowship.

%\vfill
%\newpage

\end{document}